\documentclass[twoside]{article}

\usepackage{graphicx}

\usepackage[T2A]{fontenc}
\usepackage[cp1251]{inputenc}
\usepackage{cmpj2e}

\title[Spin models of quasi-1D quantum ferrimagnets]
{Spin models of quasi-1D quantum ferrimagnets  with competing
interactions}
\author[N. B. Ivanov]{N. B. Ivanov\refaddr{label1,label2}}
\addresses{
\addr{label1} Institute of Solid State Physics, Bulgarian Academy of
Sciences, 72 Tzarigradsko chaussee, 1784 Sofia, Bulgaria
\addr{label2}  Fakult\"at f\"ur Physik, Universit\"at Bielefeld,
D-33501 Bielefeld, Germany
}
\begin{document}

\maketitle

\begin{abstract}
We present a brief survey of the recent theoretical work
related to  generic Heisenberg spin models describing 
quasi-one-dimensional quantum ferrimagnets. 
The emphasis is on  quantum chains and ladders 
with strong competing interactions,  such as the frustrated  $J_1-J_2$  
chain with  alternating (1,1/2) spins, the spin-1/2 diamond
chain with four-spin cyclic couplings, and some generic types of
mixed-spin ladders with geometric frustration. As a rule, discussed
models exhibit rich quantum phase diagrams and provide some 
interesting  examples of one-dimensional magnetic-paramagnetic quantum phase 
transitions. A number of open problems in the reviewed research area
are discussed. 

\pacs 75.10.Jm, 75.10Pq,75.45.+j
\end{abstract}

\section{Introduction}
\begin{figure}[htb]
\centerline{\includegraphics[width=0.65\textwidth]{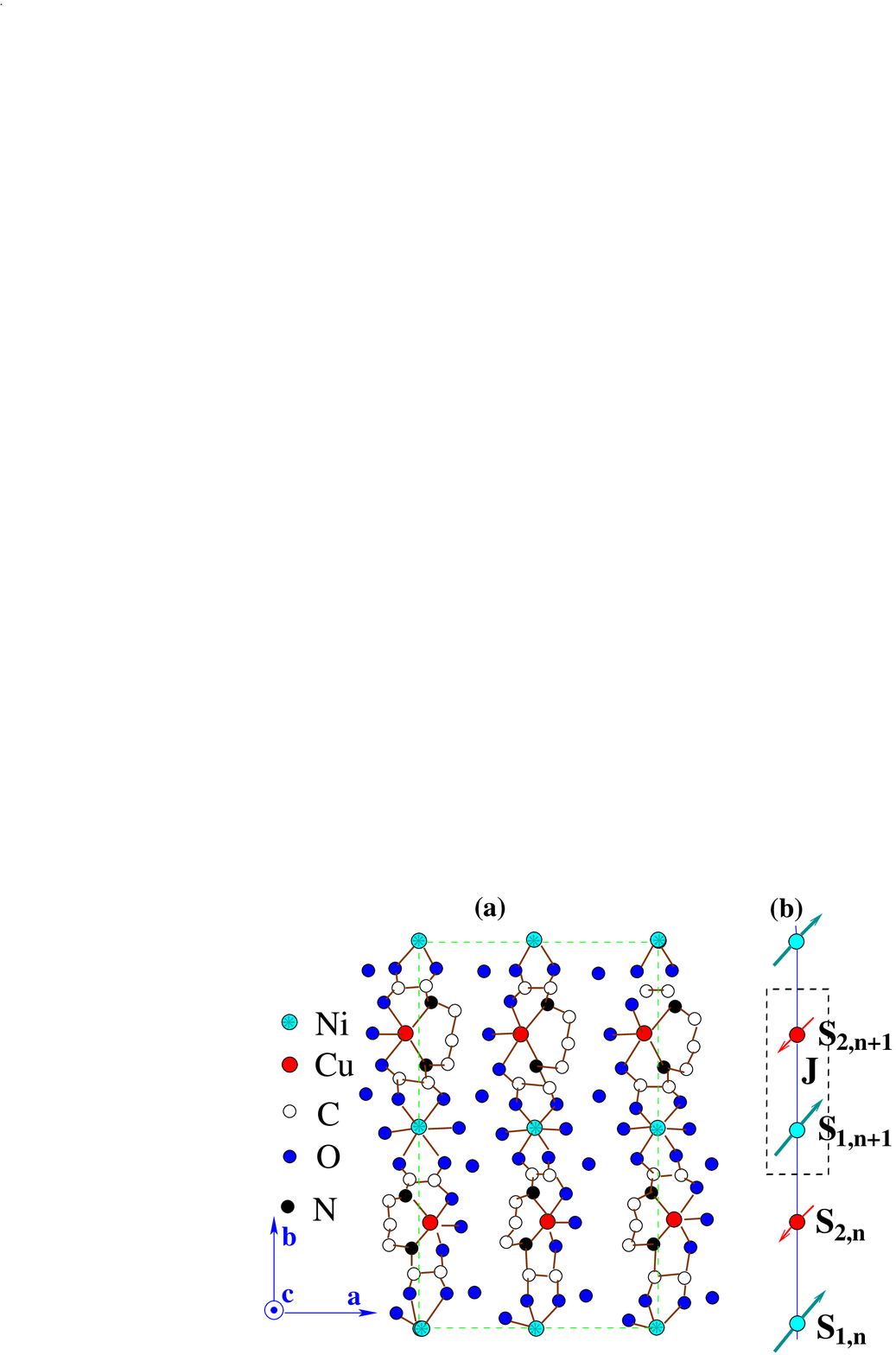}}
\caption{(a) Structure of the quasi-one-dimensional bimetallic
compound NiCu(pba)(H$_2$O)$_3\cdot$2H$_2$O
with alternating site spins
$S_1=S_{Ni}=1$ and $S_2=S_{Cu}=\frac{1}{2}$ along the axis 
$\mathbf{b}$. Hydrogen atoms are omitted for clarity \cite{hagiwara1}.
(b) The  classical N\'eel state of the mixed-spin AFM
chain model (\ref{h}) describing the bimetallic compound
NiCu(pba)(H$_2$O)$_3\cdot$2H$_2$O. 
}
\label{f1-cuni}
\end{figure}

Ferrimagnets are unsaturated magnetic materials exhibiting a net
ferromagnetic (FM)  moment, as a rule resulting  from  
magnetic sublattices with different  magnetic ions and/or 
different number of  magnetic sites \cite{wolf}. During the past two decades  
it has become  possible to synthesize a large variety of 
quasi-one-dimensional (1D) 
mixed-spin compounds with ferrimagnetic properties \cite{kahn1}. 
Most of  these materials are bimetallic  molecular magnets 
containing two different transition metal ions per unit cell, which are
alternatively distributed on the lattice (see figure~\ref{f1-cuni}). 
MnCu (dto)$_2$(H$_2$O)$_3\cdot$4.5H$_2$O (dto = dithiooxalato) 
is the first structurally characterized ferrimagnetic chain 
\cite{verdaguer,gleizes}. Two families of ferrimagnetic chains are described 
by ACu(pba)(H$_2$O)$_3\cdot$ nH$_2$O ( pba = 1,3-propylenebis)
and ACu(pbaOH)(H$_2$O)$_3\cdot$ nH$_2$O(pbaOH = 2-hydroxo-1,3 - propylenebis),
where  A=Ni, Fe, Co, and Mn \cite{pei1,kahn2,koningsbruggen}.
Cited  pioneer studies  have stimulated further
extensive  experimental research on heterometallic and 
 homometallic ferrimagnetic chains
\cite{kahn3,hagiwara1,clerac,effenberger,ohta,kikuchi}.
Another class of ferrimagnetic materials are the so-called 
\textit{topological ferrimagnets}. 
The homometallic magnetic compound 
A$_3$Cu$_3$(PO$_4$)$_4$ (A=Ca,Sr,Pb) \cite{effenberger}
is a quasi-1D example of such materials.  
In this compound, the Cu$^{2+}$ ions form   diamond chains 
with strongly coupled trimers bridged by oxygen ions. Similar magnetic 
structures appear in the magnetic
compound Cu$_3$(CO$_3$)$_2$(OH)$_2$~\cite{ohta,kikuchi} known
as \textit{azurite}. In addition, there are a number of experimental 
works on quasi-1D  organic and inorganic magnetic materials 
with similar structures and ferrimagnetic properties \cite{nishide}.
Apart from pure scientific interest, the discussed  systems
possess a potential for various technological applications.

The discussed  experimental research  has  established the  basis for
future theoretical studies on 1D quantum spin models exhibiting
quantum ferrimagnetic states. Nowadays, this is one of the areas 
in the framework of the intensive  research  on  1D magnetism  \cite{mikeska1}. 
Below we present a brief survey of the theoretical results in this hot area, 
the emphasis being on some basic  quantum spin models of quasi-1D quantum 
ferrimagnets with competing interactions. A generic spin model  of the 1D
quantum ferrimagnet is the Heisenberg spin chain with AFM  nearest-neighbor 
exchange interactions and  two different alternating spins $S_1$ and $S_2$
($S_1>S_2$). The extreme quantum  variant of this model ($S_1=1,S_2=1/2$) is
used in Section 2 to survey some basic ground-state and thermodynamic 
properties of 1D quantum ferrimagnets. The emphasis in Section 3
is on the quantum phase diagrams of  a few generic Heisenberg spin models 
describing 1D quantum ferrimagnets with strong competing interactions,
such as  the frustrated  $J_1-J_2$  chain with  alternating (1,1/2) spins, 
the spin-1/2 diamond chain with four-spin cyclic couplings, and some basic  
types of mixed-spin ladders with geometric frustration. 
The last Section contains some conclusions.
\section{ The antiferromagnetic mixed-spin Heisenberg  chain}
A generic spin model  of the 1D quantum ferrimagnet is the Heisenberg spin 
chain with AFM  nearest-neighbor exchange interaction 
($J>0$) and  alternating spins $S_1$ and $S_2$ ($S_1>S_2$), 
described by the Hamiltonian
\begin{equation}\label{h}
{\cal H}= J\sum_{n =1}^{L}\left(
{\mathbf S}_{1,n}+{\mathbf S}_{1,n+1}\right)\cdot {\bf S}_{2,n}
-\mu_BH\sum_{n =1}^{L}\left( g_1S_{1,n}^{z}
+g_2S_{2,n}^{z}\right).
\end{equation}
The integers $n$ number the unit cells, each containing two lattice 
spacings  and two kinds of quantum  spin operators ${\mathbf S}_{1,n}$
and ${\mathbf S}_{2,n}$ characterized by the  quantum spin 
numbers $S_1$ and $S_2$ ($S_1>S_2$), respectively.  
$g_1$ and $g_2$ are the $g$-factors of
the site  magnetic moments, $\mu_B$ is the Bohr magneton,  and $H$
is an  external uniform magnetic field applied along the $z$ direction.
As an example, the following  parameters of the Hamiltonian
(\ref{h}) have been extracted from magnetic measurements on the
recently synthesized quasi-1D  bimetallic  compound
NiCu(pba)(D$_2$O)$_3\cdot$2D$_2$O \cite{hagiwara1}
with a structure which is similar to the one  shown in figure~\ref{f1-cuni}:
$(S_1,S_2)\equiv (S_{Ni},S_{Cu})=(1,1/2)$,
$J/k_B=121K$, $g_1 \equiv g_{Ni}=2.22$, $g_2\equiv g_{Cu}=2.09$.
Henceforth we suggest  $g_1=g_2\equiv g$  and use the
Planck constant $\hbar$, the Boltzmann constant $k_B$,  
and the lattice spacing $a_0$ as unites. 
 
\subsection{ Ground-state properties and excitations}
\subsubsection{$H=0$}
Let us start with the case $H=0$.
According to  Lieb-Mattis' theorem \cite{lieb}, for $H=0$ the ground 
state of the bipartite model (\ref{h})  has a total spin $S_T=(S_1-S_2)L$,  $L$ 
being the  number of unit  cells, so that it is necessarily long-range ordered. 
Such ferrimagnetic ground  states  may also be referred to as 
\textit{quantized ferrimagnetic states}
since the  FM moment is quantized in integral (or half-integral)
multiples of the number of unit cells $L$. As explicitely demonstrated below
(Section 3), strong enough competing interactions can  suspend this 
quantization rule in some regions of the parameter space where the 
long-range ferrimagnetic order still survives.
Since the model has a magnetically ordered ground state, this makes the 
1D problem amenable to the spin-wave theory (SWT) approach \cite{ivanov1}. 

Valuable qualitative information about the ground state and low-lying
excitations can be extracted already from the linear spin-wave theory 
(LSWT) \cite{pati1,brehmer1,pati2,yamamoto1}.  
The on-site magnetizations $m_1=\langle S_{1,n}^z\rangle$ 
and $m_2=-\langle S_{2,n}^z\rangle$ ($m_1-m_2=S_1-S_2$) 
are parameters of the quantum ferrimagnetic phase  keeping 
information about  the long-range spin correlations. 
LSWT implies  that in the extreme quantum case $(S_1,S_2)=(1,1/2)$
the quantum spin fluctuations  reduce substantially the classical 
on-site magnetizations ($m_2/S_2\approx 0.39$). Notice, however,   
that due to the broken sublattice symmetry,  in  the mixed-spin models 
there appear  important first-order  corrections 
to $m_1$ and $m_2$ resulting from two-boson interactions in the bosonic
Hamiltonian~\cite{ivanov2}. 
Up to second order in $1/S_2$ (in respect to the linear approximation), 
SWT series  gives the precise result  $m_2=0.29388$ \cite{ivanov3}, 
to be compared with the density-matrix
renormalization-group  (DMRG) estimate $m_2=0.29248$ \cite{pati2}.
Another interesting peculiarity of the mixed-spin chain is the extremely 
small correlation length (smaller than the unit cell) of 
the  short-range spin fluctuations \cite{brehmer1,pati2}.  
This observation explains the good quantitative description of the model 
achieved  with the variational approach using matrix-product states 
\cite{kolezhuk1,kolezhuk2}.     
\begin{figure}[htb]
\centerline{\includegraphics[width=0.65\textwidth]{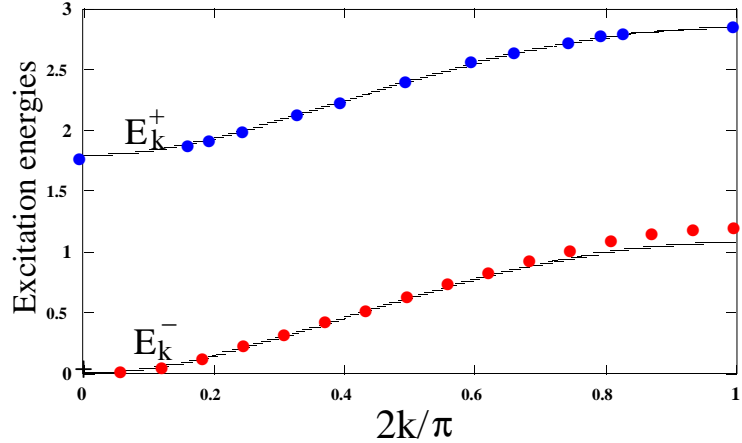}}
\caption{Dispersions of the one-magnon excitations 
($E_k^{\pm}$) in the system $(S_1,S_2)=(1,1/2)$ calculated up to second order   
in the $1/S_2$ relative to the linear  approximation~\cite{ivanov4}. 
The points in  $E_k^{\pm}$ show numerical 
ED results for periodic chains  (for $E_k^{+}$)  
and QMC results (for $E_k^{-}$)~\cite{yamamoto1}}.
\label{en}
\end{figure}

Turning to the excitation spectrum,  SWT predicts
two types of low-lying magnon excitations 
$E_k^{\pm}$, respectively, in the subspaces with   z components of the 
total spin $S_T^z= S_T\pm1$ (see figure~\ref{en}). 
In the long wavelength limit $k\ll 1$, the
FM  mode $E_k^{-}$ takes the  Landau-Lifshitz form
\begin{equation}\label{ef}
E_k^{-}=\frac{\varrho_s}{M_0}k^2
+{\cal O}(k^4)\, ,
\end{equation}
where  $\varrho_s=JS_1S_2$ is the spin stiffness constant
\cite{halperin}
and $M_0=(S_1-S_2)/2$ is the linear density of the net FM 
moment. This form of the Goldstone modes is typical for
Heisenberg ferromagnets,  and reflects the fact that the FM 
order parameter is itself a constant of the motion.  As demonstrated below,
the parameters $\varrho_s$ and $M_0$ 
play a basic role in the low-temperature thermodynamics 
of the model. On the other hand, the AFM branch 
$E_k^{+}$ is gapped, 
 with a minimum at $k=0$ given by  $\Delta^{+}=2(S_1-S_1)J$ in a LSWT
approximation. Note, however,  that the  LSWT estimates for 
$\varrho_s$ and the AFM gap $\Delta^{+}$  
are further  renormalized by the boson-boson 
interactions \cite{ivanov4}: In the extreme quantum system  
$(S_1,S_2)=(1,1/2)$,  already the first order corrections in $1/S_2$ 
give the results  $\varrho_s /(Js_1s_2)=0.761$ 
and $\Delta^{(+)} =1.676J$. The  quantum Monte-Carlo (QMC)
result for this parameter  $1.759$ \cite{yamamoto1} clearly 
indicates  the importance of the $1/S_2$ corrections. 
Precise estimates for the parameters of the $(1,1/2)$ chain 
have also been obtained by using cluster series expansions~\cite{zheng}:
$m_2=0.292487(6)$, $\Delta^+/J=1.7591(6)$ and 
$\varrho_s /(Js_1s_2)=0.831(5)$.
\subsubsection{$H\neq 0$}
The introduction of a Zeeman term in Eq.~(\ref{h})
leaves all eigenstates with a given  z component of the total spin
 $S^{z}_{T}$ invariant, while shifting the eigenenergies
by $hS^{z}_{T}$, where $h=g\mu_BH$. 
In particular,  the one-magnon states  $E_k^{\pm}$,  carrying magnetizations 
$\pm 1$, change to $E_k^{\pm}(h)=E_k^{\pm} \mp h$, i.e, 
 the Zeeman term introduces a gap $\Delta^{-}(h)=h$ for  the
FM branch and reduces the gap of the AFM
branch, $\Delta^{+}(h) = \Delta^{+} - h$.
If the magnetic field exceeds the critical value $h=h_{c1}= \Delta^{+}$,
the AFM  gap closes and the system enters a critical phase 
which  is expected to be a kind of  Luttinger liquid, 
in analogy with the behavior of other gapped spin models like the spin-1 
chain and the spin-1/2 ladder~\cite{affleck1,giamarchi1,giamarchi2}.
The critical phase terminates at a  second critical field $h=h_{c2}= 3J$ 
at which the system becomes fully polarized~\cite{maisinger}. An accurate
description of the mixed-spin ($1,1/2$) Heisenberg chain  ($H\neq 0)$
in the critical phase $h_{c1}<h<  h_{c1}$ has been achieved by
a mapping to an effective  spin-1/2 XXZ chain in external magnetic
field which uses  variational  matrix-product states~\cite{kolezhuk2}.
Interestingly, such a  critical phase seems to be characteristic for the 
entire class of mixed-spin $(S_1,S_2)$ anisotropic  XXZ Heisenberg chains
 with easy-plane anisotropy and short-range interactions.  
An extensive  numerical analysis, using exact-diagonalization (ED)
 results for $(1,1/2)$ and  $(3/2,1/2)$  periodic chains, suggests universal 
critical properties for these  chains in the entire interval from the FM
to the ferrimagnetic isotropic points~\cite{alcaraz1}. 
Along this  phase, the critical fluctuations are ruled by a conformal 
field theory of Gaussian type with a topological charge $c=1$.
  
Discussed one-magnon excitations control the magnetization
process of the mixed-spin chain. Since the lowest excitation
increasing the FM moment 
$M=\langle S^z_{1,n}+S^z_{2,n}\rangle$ exhibits an energy gap, 
the magnetization curve $M(H)$ has a plateau at $M=S_1-S_2$. This
plateau phase, however, appears in the related classical system, as well. 
In this connection, it is interesting to examine the role of different  
anisotropies which may appear in real materials. As to the  exchange 
anisotropy, it occurred that the quantum effects simply stabilize the 
plateau at  $M=S_1-S_2$ against the XY anisotropy, i.e., 
the effect is reduced to quantitative changes of the classical 
result~\cite{sakai1}. A clear indication for the quantum origin of the  
discussed magnetization plateau was provided for the $(1,1/2)$ chain 
with a single-ion anisotropy for  spin-1 sites, 
$ D\left( S^z_{1,n}\right)^2$, which is even more realistic  for the 
existing mixed-spin  materials~\cite{sakai2}. 
It was revealed, in particular,  that the mechanism
of the $M=1/2$ plateau can change from a Haldane to a large-$D$ type 
trough a Gaussian quantum critical point, estimated as $D/J=1.114$, which 
is a kind of justification for the quantum origin of the $M=1/2$
plateau in this system. Two  plateau phases  ($M=1/2$) with  
different parities  have also been studied in a  frustrated mixed-spin 
chain with different next-nearest-neighbor AFM 
bonds (see figure~\ref{models}a)~\cite{kuramoto1}.  
\subsection{Thermodynamics}
\begin{figure}[htp]
\centering
\begin{minipage}[c]{0.4\textwidth}
\centering
\includegraphics[width=1\textwidth]{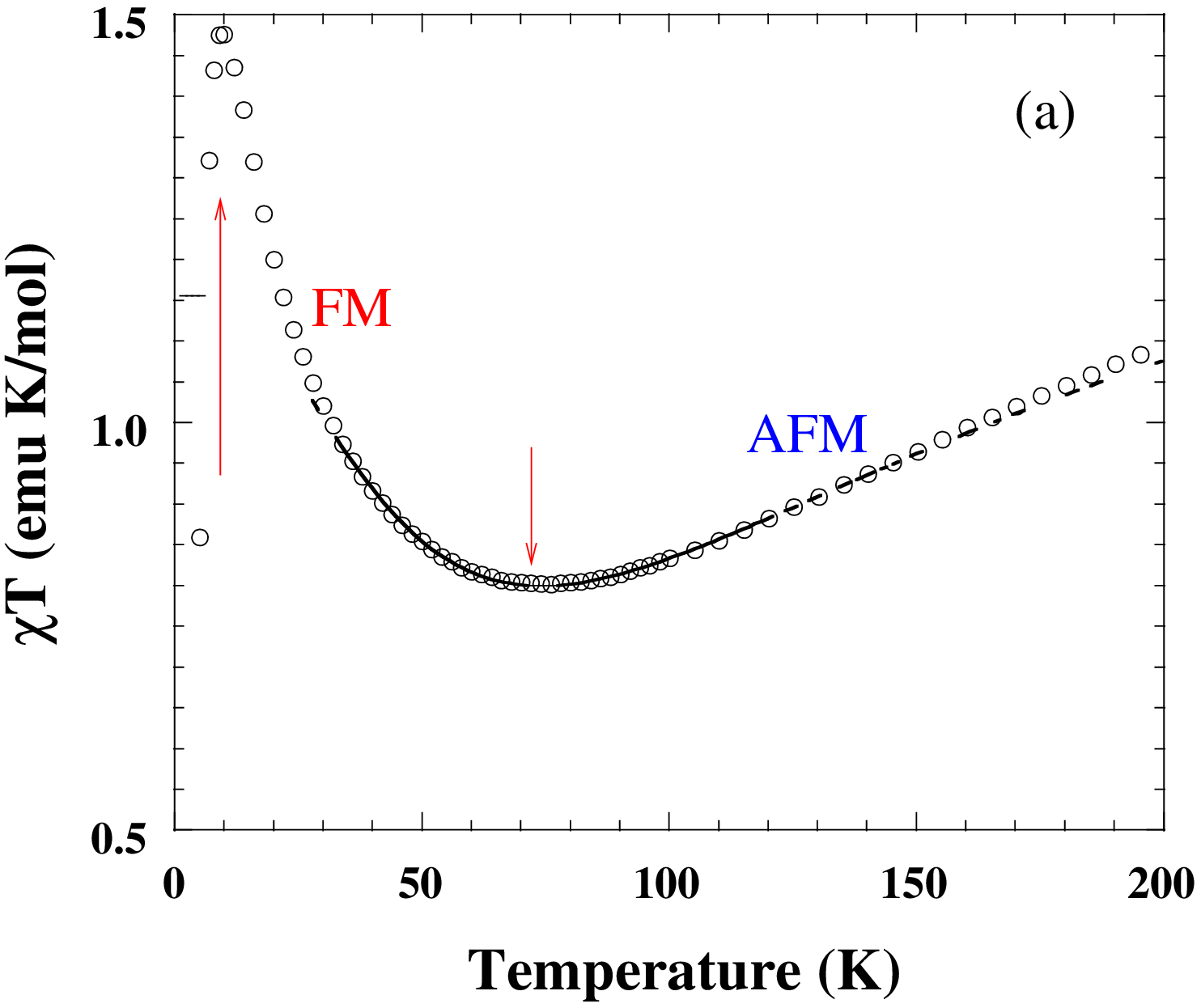}
\end{minipage}
\hspace{0.3in}
\begin{minipage}[c]{0.4\textwidth}
\centering
\includegraphics[width=1\textwidth]{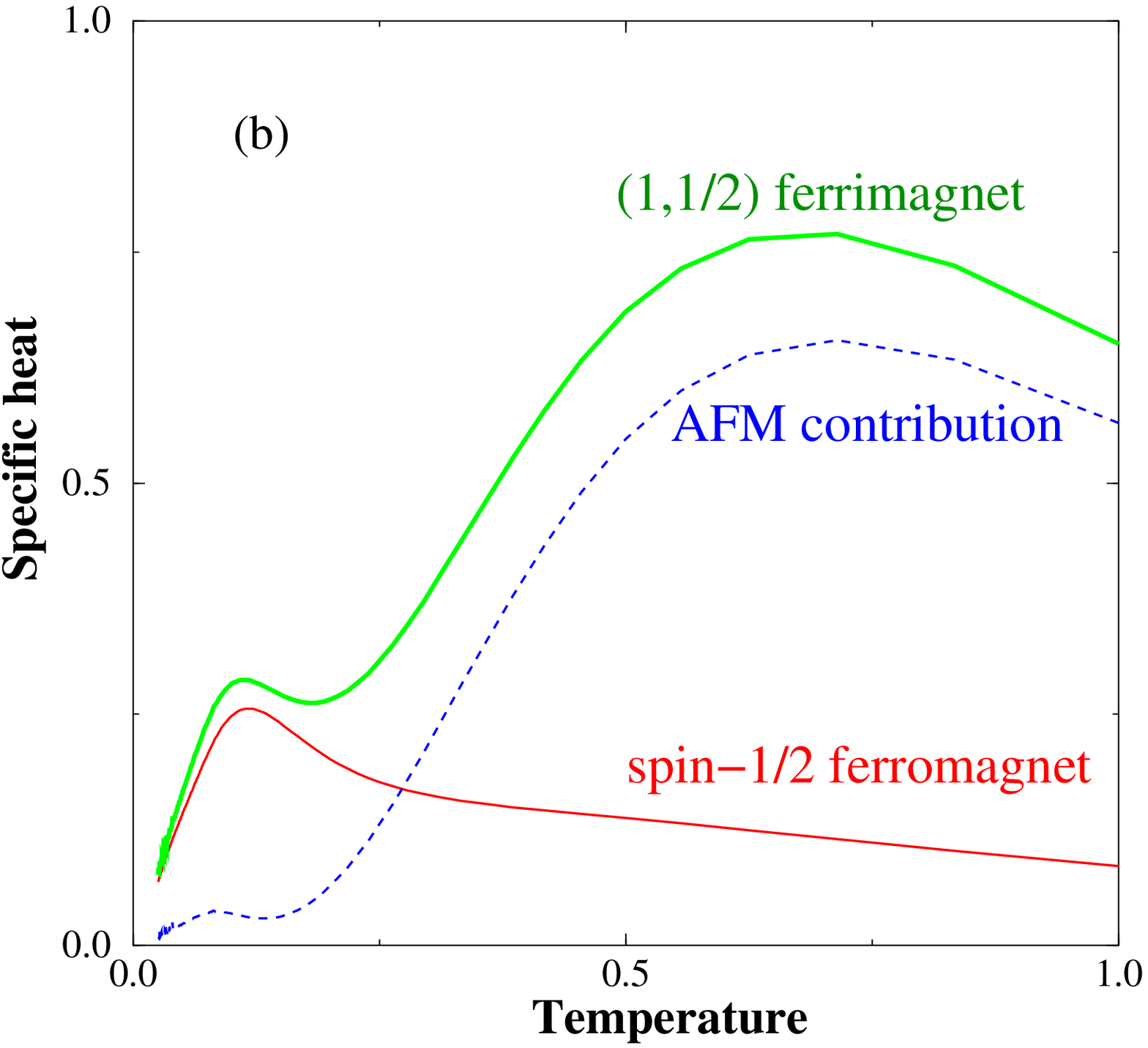}
\end{minipage}
\caption{(a) Experimental results ($\chi\cdot T$ \textit{vs} $T$)
for the $(1,1/2)$ mixed-spin chain
NiCu(pba)D$_2$O)$_3\cdot$ 2D$_2$O~\cite{hagiwara1}.
(b) $C_v$ \textit{vs} $T$ of the $(1/1/2)$ mixed-spin chain
(DMRG) compared to $C_v(T)$ of the spin-1/2 FM chain (DMRG) at
$h=0.05J$.  The difference of both specific heats 
can be identified as AFM contribution~\cite{maisinger}.  
$C_v$ and $T$ are presented in the unites of $J$. 
}
\label{thermodynamics}
\end{figure}

Most of the  experiments on quasi-1D mixed-spin systems  carried out in the past
concerned the  magnetic susceptibility of ferrimagnetic chains with
rather big $S_1$ spins (typically, $S_1=5/2)$), making the system  
rather classical~\cite{kahn4}. As a matter of fact, quantum effects 
are most pronounced in the extreme quantum case 
$(S_1,S_2)=(1,1/2)$,  realized, for example,  in the quasi-1D  
bimetallic material  NiCu(pba)D$_2$O)$_3\cdot$ 2D$_2$O~\cite{hagiwara1}. 
As mentioned above, this material has an exchange interaction of about 
121 K, and shows a three dimensional AFM ordering transition at 7 K
(see figure~\ref{thermodynamics}) which somewhat obscures  the very low 
temperature  ($T$) features of the 1D ferrimagnet. 
Still, low-field measurements  are in good agreement  with the 
theoretically expected  results. On the other hand, the magnetic
fields necessary to reach the theoretically most interesting critical  phase 
would be for this compound well beyond 100 T, making the search for other  
compounds with weaker interactions an  interesting challenge for experimentalists.

Low-temperature thermodynamics ($T\ll \varrho_sM_0$) of the  mixed-spin
chains is controlled by the FM magnons, so that  in zero magnetic field 
it can be described by using Takahashi's constrained SWT for 
ferromagnets~\cite{takahashi1,takahashi2}. A few variants
of this theory have also been applied to 1D  
ferrimagnets~\cite{yamamoto2,yamamoto3,wu}. In terms of the ground-state
parameters $\varrho_s$ and $M_0$, the explicite form of the series in 
powers of $t\equiv T/\varrho_sM_0$ for the uniform susceptibility
$\chi$ and the specific heat $C_v$ read as~\cite{ivanov5}     

\begin{eqnarray}\label{cv}
   \frac{\chi T}{M_0}
      &=& \frac{2}{3}t^{-1}
      - \frac{\zeta(\frac{1}{2})}{\sqrt{\pi}}t^{-\frac{1}{2}}
      +\frac{\zeta^2(\frac{1}{2})}
      {2\pi}+{\cal O}\left( t^{\frac{1}{2}}\right)
         \,,\nonumber \\
   \frac{C_v}{M_0}
      &=& \frac{3}{8}\frac{\zeta(\frac{3}{2})}{\sqrt{\pi}}
         t^{\frac{1}{2}}-\frac{t}{2} +\left[const
         -\frac{15\zeta(\frac{1}{2})}{32\sqrt{\pi}}\right]
t^{\frac{3}{2}}+{\cal O}\left(t^2\right).
\end{eqnarray}
Here,
$\zeta(z)$ is Riemann's zeta function, and
$const=15(S_1^2+S_1S_2+S_2^2)\zeta(\frac{5}{2})/128\sqrt{\pi}$.

As may be expected, the above expansions reproduce Takahashi's original 
expansions for the Heisenberg FM chain characterized by the parameters 
$M_0=S$ and $\varrho_s=JS^2$.  Apart from the coefficient of  $t^{3/2}$
in the expansion  for $C_v$, the above expressions
reproduce precisely the thermodynamic Bethe-ansatz calculations for 
the spin-1/2 FM Heisenberg chain~\cite{takahashi3,yamada}.
It is interesting to note that without the factor $const$ in the expression
for $C_v$, both expansions fulfill the general hypothesis according to which
 in 1D Heisenberg ferromagnets all observables should be universal
functions of the bare couplings $M_0$, $\varrho_s$, and $h$,
realizing a {\em no-scale-factor universality}~\cite{read,takahashi4}.
In particular, this means that  the free-energy density
should have  the generic form
\begin{equation}
\frac{F}{N}=TM_0\Phi_F \left( \frac{\rho_sM_0}{T},\frac{h}{T}\right)\, ,
\end{equation}
where $\Phi_F(x,y)$ is a universal scaling function with no
arbitrary scale factors: for example, the site spin $S$ enters
only indirectly, through $\varrho_s$ and $M_0$. Appearance of
the  factor violating the above scaling hypothesis
is probably an artifact of Takahashi's SWT. As far as we know, 
up to now universal low-$T$ thermodynamic properties of  
quasi-1D quantum ferrimagnets have not been studied.  

 Whereas low-$T$ thermodynamics
reflects the dispersion relation of the lower gapless
FM branch  $E_k^-$,  the intermediate-$T$
behavior is dominated by that of the upper  AFM branch
$E_k^+$. Therefore, when the condition $T\ll \Delta^+$ is violated,
all the thermodynamic parameters are expected to demonstrate a
crossover from a FM to a AFM behavior. In particular,
for the uniform susceptibility $\chi$ one may expect 
$\chi(T)\propto T^{-2}$  (as $T \rightarrow 0$) and  
the Curie law behavior $\chi(T)\propto T^{-1}$
in the  high-$T$ region.  Figure~\ref{thermodynamics}a displays  a 
typical  experimental curve ($\chi T$ \textit{vs} $T$) which is
usually used to determine  the character of the short-range interactions. 
For a  paramagnetic system, Curie's law implies the behavior 
$\chi T =const$  over the whole $T$ range. If the magnetic system has 
dominant  FM (AFM) interactions,  $\chi T$ increases (decreases) 
when $T$ is decreased. Thus, in the interval  $70K<T<300K$ the dominating 
coupling between 
nearest-neighbor spins is AFM, whereas the increase of  $\chi T$ below 
the minimum around $T\approx 70$ K implies that the system behaves like a 
FM chain at low $T$. The steep decrease below $10$ K may be attributed to 
an established 3D AFM long-range order.  

Turning to the behavior of the specific heat, it changes from
$C_v(T)\propto T^{1/2}$ in the extreme low-$T$ region [see Eq.~(\ref{cv})],
through  an AFM  Schottky-like peak (for intermediate $T$) to 
a paramagnetic  $T^{-2}$ decay for high $T$ 
(see figure~\ref{thermodynamics}b). In addition, $C_v(T)$  
acquires  a characteristic double-peak structure 
for $0<h<h_{c1}$,  which is  related to
the gapped modes  $E_k^{\pm}$~\cite{yamamoto2}. As a matter of fact, both 
gapped antiferromagnets as well as ferromagnets in an external magnetic field 
exhibit a pronounced peak whose position is related to the gaps 
$\Delta^{\pm}$. Such a double-peak structure has also been predicted 
for the  $(1,1/2)$ chain with a FM exchange constant 
($J<0$)~\cite{fukushima}.    
\section{ Mixed-spin chains and ladders with magnetic frustration }
 Over the last years there has been an increasing interest  
in  quantum  spin systems  with competing  exchange interactions
\cite{conf}. Quasi-1D quantum  spin systems  with geometric 
frustration, both for half-integer and integer  spins
\cite{bursill,chitra,scholl,kolezhuk3,wite1} 
(see also the review article~\cite{mikeska1}) set up  an important
part of this research. It is remarkable that up to now a relatively small 
amount of research on 1D frustrated mixed-spin models has been published.
These models typically exhibit ground states with a net FM moment,
so that a number of intriguing issues like  the nature of the new phases
as well as  the character of the ferrimagnetic-paramagnetic transitions
may be studied.  What makes such transitions interesting
is the fact that the order parameter is  a conserved quantity:
As is known, in systems with a Heisenberg $SO(3)$ symmetry this 
conservation law is  expected to lead to strong constraints on the critical
field theories~\cite{sachdev1,sachdev2}.
 Below we survey the available results for some generic  
Heisenberg spin models relevant for quasi-1D quantum ferrimagnets 
with competing interactions. 
\begin{figure}[htb]
\centerline{\includegraphics[width=0.65\textwidth]{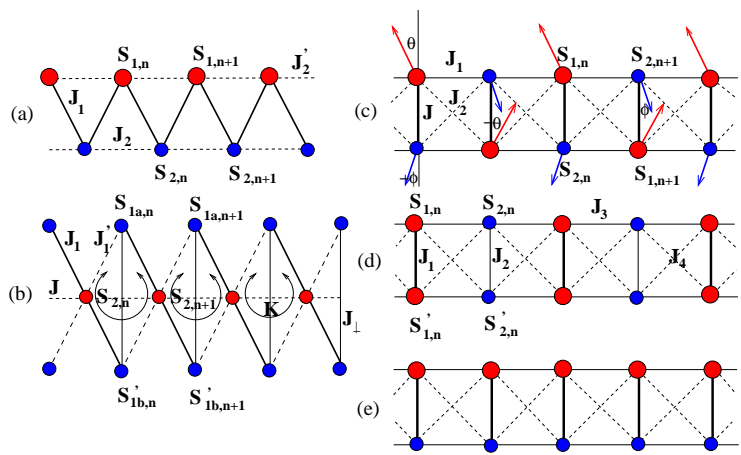}}
\caption{Some generic  1D  Heisenberg spin  models 
with AFM exchange bonds exhibiting ferrimagnetic ground states: 
(a) The mixed-spin $J_1-J_2$ chain  with AFM nearest- ($J_1$) and 
next-nearest-neighbor ($J_2,J_2^{'}$) exchange bonds; 
(b) The distorted ($J_1\neq J_1^{'}$)  diamond chain 
with frustrating vertical bonds ($J^{'}$) and  
four-spin cyclic exchange interactions ($K$); 
(c,d,e) Three generic types of mixed-spin ladders with geometric frustration. 
}
\label{models}
\end{figure}

\subsection{Mixed-spin $J_1-J_2$ Heisenberg chains}
The mixed-spin Heisenberg chain composed of two types of alternating spins  
($S_1>S_2$),  which are connected  via  competing nearest-neighbor ($J_1>0$) and 
next-nearest-neighbor ($J_2, J_2^{'}>0$) AFM exchange bonds 
(see figure~\ref{models}a), is one of the simplest realistic models of 
frustrated 1D quantum ferrimagnets~\cite{ivanov2,kuramoto1}. 
This model may also be  considered as a ferrimagnetic analogue  of  
the frustrated  FM Heisenberg chain with  FM nearest-neighbor- ($J_1<0$) 
and AFM  next-nearest-neighbor ($J_2>0$) exchange bonds. 
The  spin-1/2 frustrated $J_1-J_2$ FM chain  has recently attracted 
much attention, as it is supposed to describe a number of quasi-1D
edge-sharing cuprates, such as Rb$_2$Cu$_2$Mo$_3$O$_{12}$~\cite{hase} 
and LiCuVO$_4$~\cite{enderle}. The latter material exhibits  
multiferroic properties~\cite{naito} as well as an interesting 
 specific  phase transition in a magnetic field from an ordered  spiral  
to an ordered  modulated-collinear  magnetic phase~\cite{banks}. 
On the theoretical side, the latter FM model was shown to exhibit a  
vector chiral long-range  oder (in a moderate magnetic field) 
as well as a rich variety of exotic quantum phases with different kinds of 
multipolar spin correlations
(in a larger magnetic field)~\cite{hikihara,sudan}.             
   
Turning to the mixed-spin $J_1-J_2$ chain, let us begin with a comment
on its classical limit. The ground state of the model can be described
by the ansatz  $\mathbf{S}_{i,n}=S_i\left[ \mathbf{u} \cos(Qn)+
\mathbf{v}\sin(Qn)\right]$ ($i=1, 2$), where  $\mathbf{u} \perp \mathbf{v}$ 
are unit vectors in the spin space. The classical  ferrimagnetic 
state with a pitch angle between neighboring spins $Q/2=\pi$  
is stable up to the phase transition  point 
$J_{2c}=S_1S_2J_1/[2(S_1^2+S_2^2)]$. In the strongly frustrated region 
$J_2>J_{2c}$, the stable state is a spiral with an ordering wave vector given by
$\cos (Q/2)=-S_1S_2J_1/[2J_2 (S_1^2+S_2^2)]$.
In the limit $J_2\rightarrow\infty$, $Q=\pi$ and the system is composed of 
two decoupled  AFM chains with site spins $S_1$ and $S_2$.

Next, let us discuss the extreme quantum case $(S_1,S_2)=(1,1/2)$. 
Already a qualitative  semiclassical  analysis implies  that near 
the classical transition point at $J_2=0.2J_1$ the FM magnon branch
$E_k^{-}$  (see figure~\ref{en}) is strongly  flattened, whereas  the optical 
AFM branch shows only a smooth increase of the AFM gap $\Delta^+$. 
This means that the optical  magnons do not play any important role in the 
mechanism of the  transition. On approaching the classical transition point,
the frustration and quantum spin fluctuations play somewhat different roles:
Whereas the magnetic frustration strongly reduces the 
short-range  spin  correlation length in the ferrimagnetic phase, the 
quantum fluctuations stabilize the ferrimagnetic order up to the
point  $J_2=0.231J_1$ beyond the classical transition point at
$J_2=0.2J_1$. Another effect of the quantum fluctuations is related 
to the change of the character of the classical transition: 
A detailed  DMRG analysis of the low-energy  levels around the  
classical transition clearly indicates a level crossing, i.e., 
a first-order quantum phase transition to a singlet  ground state at  
$J_2=0.231J_1$. Further, DMRG  has also indicated
that at least from $J_2=0.25J_1$ upwards the discussed  model exhibits a 
singlet ground state~\cite{ivanov2}. 

A  number  of open issues related to the mixed-spin $J_1-J_2$
Heisenberg model can be indicated. One of the important open problems
concerns   the nature of the singlet ground  states  established  beyond 
the ferrimagnetic phase.  Since in 1D systems the classically broken 
SO(3) symmetry in the spiral state is generally expected 
to be restored by quantum fluctuations, one is enforced 
to look for possible magnetically disordered states. 
A valuable  information can be obtained from the Lieb-Schultz-Mattis 
theorem~\cite{lieb2} adapted to mixed-chain spin models~\cite{fukui1}.
The theorem is applicable to systems with a half-integer cell spin 
($S_1+S_2$) and says that  the model either has gapless
excitations or else has degenerate ground states. Therefore, one may look 
for phases with presumably some broken discrete symmetry. 
In particular, the long-ranged chiral phase found in the
frustrated spin-1/2 FM $J_1-J_2$ model in the small magnetic field region
could be a possible candidate~\cite{hikihara}.  However,
the variety of possible non-magnetic quantum phases may be 
considerably  enlarged due to the existence of two kinds of site
spins in the mixed variant of the parent model.

\subsection{Frustrated diamond  chains}
\begin{figure}[htp]
\centering
\begin{minipage}[c]{0.4\textwidth}
\centering
\includegraphics[width=1\textwidth]{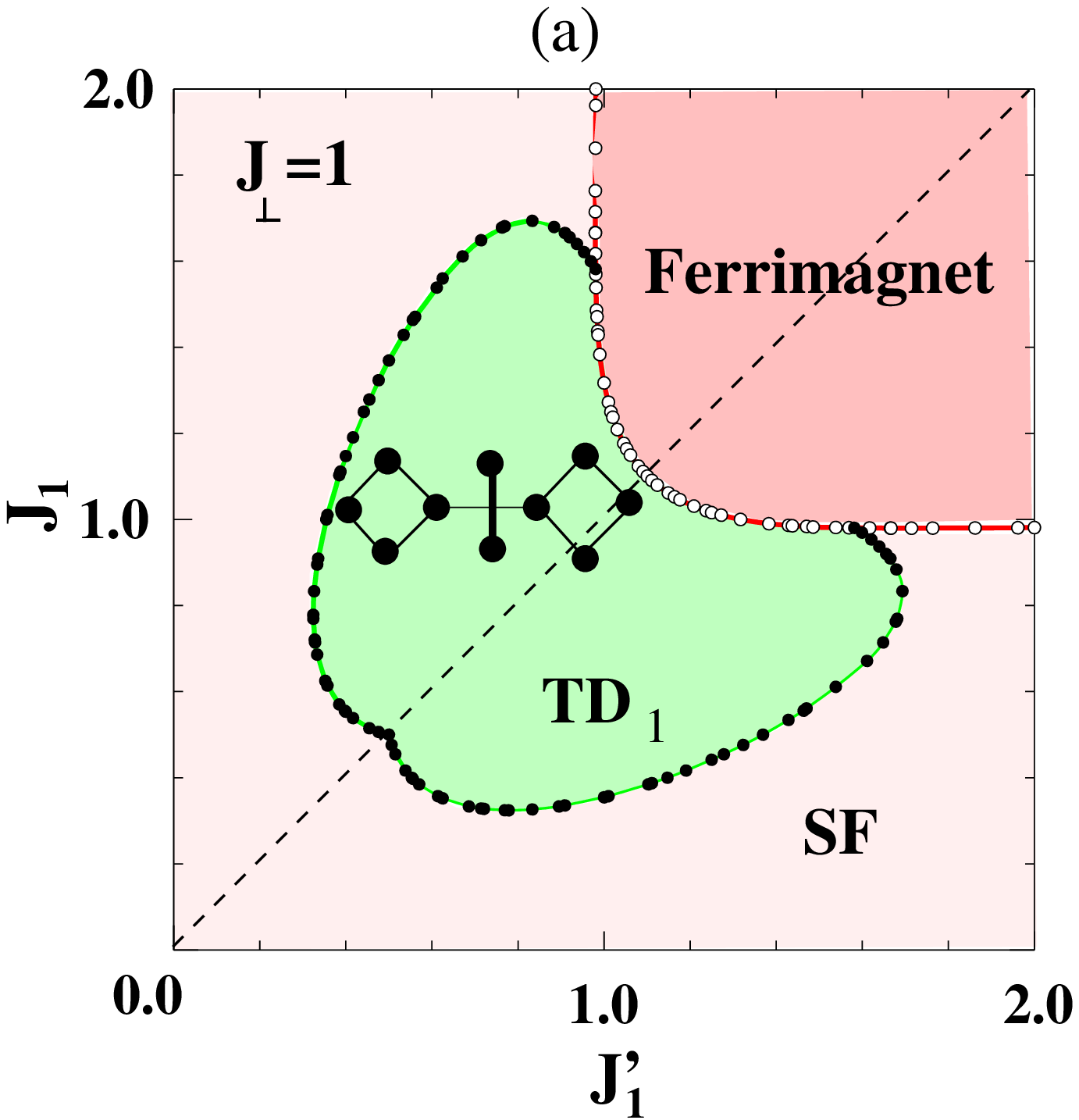}
\end{minipage}
\hspace{0.3in}
\begin{minipage}[c]{0.4\textwidth}
\centering
\includegraphics[width=0.9\textwidth]{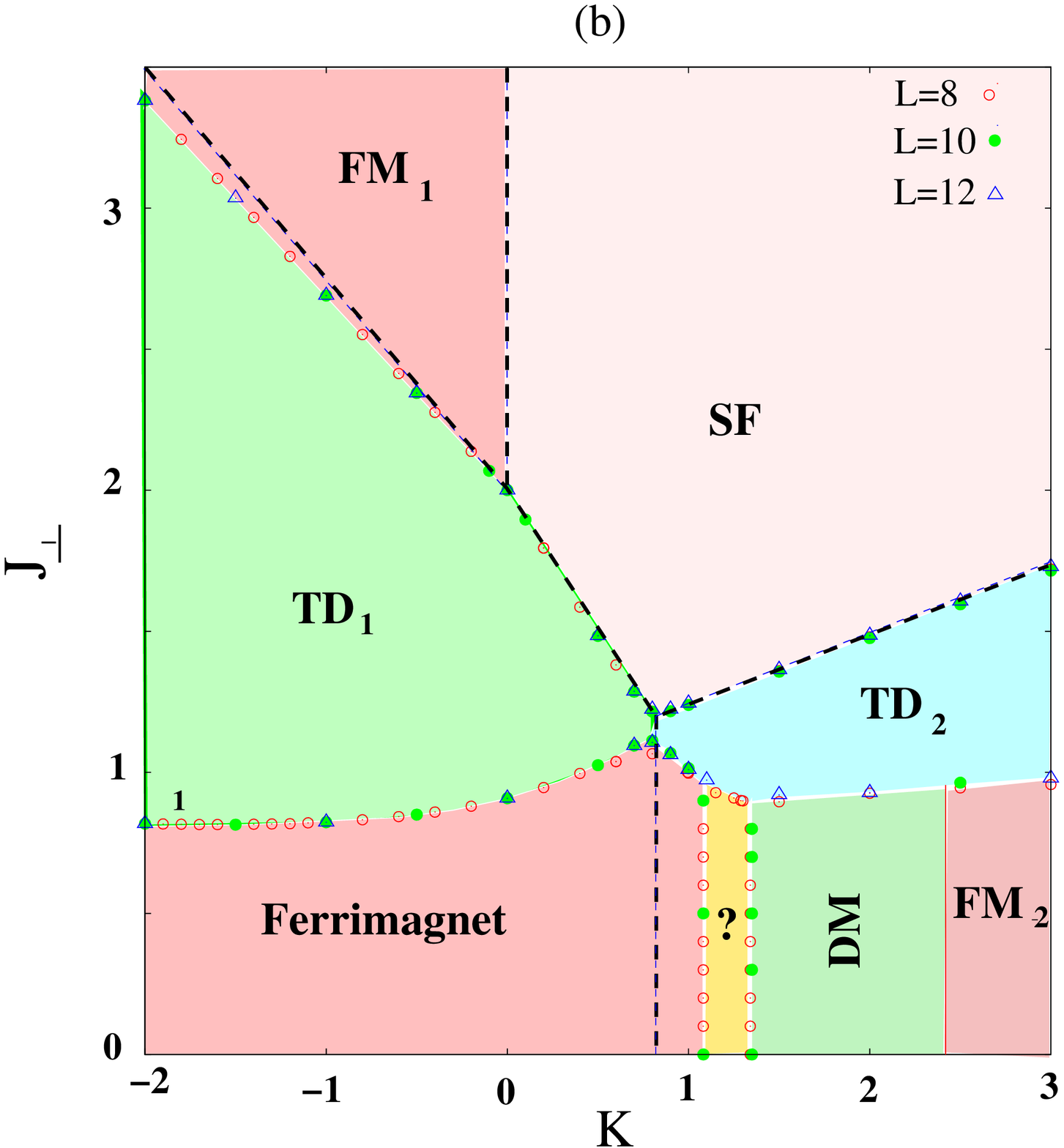}
\end{minipage}
\caption{(a) Quantum phase diagram ($T=0$) of
the distorted diamond chain in the parameter space $J_1$ \textit{vs.}
$J_1^{'}$
($J_{\perp}=1$)~\cite{tonegawa}. SF indicates the gapless spin-fluid
phase, whereas TD$_1$ marks the tetramer-dimer phase constructed
from alternating dimers and tetramers in local singlet states. (b)  
Quantum phase diagram ($T=0$) of a frustrated SDC with competing four-spin
cyclic interactions in the parameter space  $J_{\perp}$  \textit{vs.} $K$
($J_1^{'}=J_1^{'}=1$)~\cite{ivanov6}. In this system, there appear at least
five new phases denoted as follows: FM$_1$ (fully-polarized FM phase), FM$_2$
(another phase with a net FM moment),  DM (fourfold degenerate singlet
phase), TD$_2$ (another tetramer-dimer phase with tetramers in local triplet
states), and an exotic  non-Lieb-Mattis ferrimagnetic state  
denoted by a question mark. The dashed lines trace the phase boundaries
of the single-diamond system.
}
\label{phases}
\end{figure}

  The diamond Heisenberg chain is another  generic model  
of 1D quantum ferrimagnets constructed from one kind of
spins living on sublattices with  different number of sites 
(see figure~\ref{models}b). The frustrated symmetric diamond chain
(SDC) ($J_1=J_1^{'}$, $J=0$) with  AFM vertical bonds $J_{\perp}>0$  
is probably the first studied  model of  1D quantum ferrimagnet  
with competing interactions~\cite{takano1,niggemann}.
A particular variant of the  frustrated model, the distorted 
($J_1\neq J_1^{'}$) spin-1/2 diamond chain, 
has received increasing theoretical~\cite{okamoto1,mikeska2} 
as well as  experimental interest in the past decade due to its rich 
quantum phase diagram~\cite{tonegawa} (see figure~\ref{phases}a) 
and the relevance for the real material Cu$_3$(CO$_3$)$_2$(OH)$_2$ 
(azurite)~\cite{rule}. Without external  magnetic
field, three quantum phases in the parameter space
($J_1/J_{\perp},J_1^{'}/J_{\perp}$) have been discussed for the distorted
model: For $J_1/J_{\perp},J_1^{'}/J_{\perp}\ll 1$, the low-energy sector is
governed by an effective spin-1/2 AFM Heisenberg model, which indicates the
formation of a gapless spin-fluid phase (SF) with some additional 
high-energy
modes related to local excitations of the vertical  dimers.
For intermediate $J_1/J_{\perp}$ and $J_1^{'}/J_{\perp}$, the ground state
dimerizes, forming a twofold degenerate sequence of alternating tetramers
and dimers (TD$_1$ phase). Finally, for both  $J_1/J_{\perp}$ and
$J_1^{'}/J_{\perp}$ sufficiently large, the ground state is a 1D
ferrimagnet. These phases can be clearly identified already in the SDC:
SF ($J_1<0.5 J_{\perp}$), TD$_1$ ($<0.5 J_{\perp}<J_1<1.10  J_{\perp}$ ),
and 1D ferrimagnet ($J_1 > 1.10  J_{\perp}$).

The diamond Heisenberg  chain is also one of the simplest quantum spin models 
admitting a four-spin cyclic exchange coupling.  Below we discuss the impact 
 of this competing interaction on the  quantum phase diagram ($T=0$) of 
the frustrated SDC~\cite{ivanov6}. The schematic Hamiltonian of the
model is presented in figure~\ref{models}b, the standard 
 four-spin cyclic exchange coupling~\cite{muller} in a single diamond
being controlled by the parameter $K$. It is important to notice that
the cyclic coupling  does not violate the local symmetry of the frustrated 
($J_{\perp}>0$) SDC model under the exchange of  pairs  of off-chain spins  
$\mathbf{S}_{1a,n}$  and $\mathbf{S}_{1b,n}$ in the  diamonds. 
Thus, in the important case of spin-$1/2$ off-chain
operators,  the system is characterized (as in the frustrated model without 
cyclic interactions) by $L$  local
good quantum numbers $s_n=0,1$ ($n=1,2,\dots,L$) related to the composite
off-chain spins  $\mathbf{S}_{1,n}=\mathbf{S}_{1a,n}+\mathbf{S}_{1b,n}$: 
$\mathbf{S}_{1,n}^2=s_n(s_n+1)$. Using this local symmetry, the  Hamiltonian 
of the SDC can be represented in the compact form
\begin{equation}\label{hc}
{\cal H}_c=E_0+\sum_{n=1}^L\left[
J_1\mathbf{S}_{1,n}\cdot (\mathbf{S}_{2,n}+\mathbf{S}_{2,n+1})
+J_n\mathbf{S}_{2,n}\cdot \mathbf{S}_{2,n+1}
+\frac{K}{2}\{\mathbf{S}_{1,n}\cdot\mathbf{S}_{2,n},
\mathbf{S}_{1,n}\cdot\mathbf{S}_{2,n+1}\}\right].
\end{equation}
Here $E_0=J_{\perp}\sum_{n=1}^L[s_n(s_n+1)-3/4]$ is a fixed
number for every sector defined as a sequence of the local quantum numbers 
$\left[ s_1,s_2,\dots,s_L\right]$, $J_n=J +K/4 - s_nK$, and $\{ A,B \}$ is 
the anticommutator of the operators $A$ and $B$.  

Let us briefly comment the phase diagram for the spin-1/2 SDC 
(figure~\ref{phases}b), by using some symmetries of the Hamiltonian 
(\ref{hc}). In the parameter space where the ground state  
is characterized by $s_n=0$ ($n=1,2,\ldots ,L$), the first and third terms in 
the square brackets vanish and the   model is equivalent to the 
spin-$1/2$ Heisenberg chain with an effective exchange parameter $J+K/4$.
This explains the presence of a fully polarized  FM$_1$ ($J+K/4<0$) and a 
gapless spin-fluid ($J+K/4>0$) phases for large enough $J_{\perp}/J_1$.  
Note that besides the well-documented collective modes, these  phases 
 exhibit specific additional   single-particle   modes which are related  to 
local triplet excitations on the vertical dimers.
Being eigenstates of the Hamiltonian, these excitations are completely localized
in the SDC.   

For intermediate values of the parameter $J_{\perp}/J_1$, 
the ground state lies in the sector $[1,0,\cdots ,1,0]$.
The tetramer-dimer phase TD$_1$ studied in the DDC model (see
figure~\ref{phases}a) survives in the region with relatively small
cyclic interactions. This doubly degenerate singlet state  may roughly be
thought of as  a product of plaquette singlet states 
on every second diamond, as depicted in  figure~\ref{phases}a.
For larger $K$  at fixed $J_{\perp}$, there appears  another tetramer-dimer 
pase (TD$_2$), with  every  second diamond approximately  in a triplet
state. The numerical ED analysis shows relatively strong AFM correlations 
between the neighboring triplet  diamonds, as opposed to the TD$_1$ state
where the  diamonds are weakly correlated. Clearly, both  tetramer-dimer
phases are gapped and doubly-degenerated.

Finally, for  $J_{\perp}/J_1<1$ the ground state lies in  the sector
$s_n=1$ ($n=1,2,\ldots ,L$), so that the low-energy region of  Eq.~(\ref{hc})
describes a mixed-spin $(1,1/2)$ chain with  competing three-spin exchange
interactions. To the best of our knowledge,  realistic 1D mixed-spin  
Heisenberg models with multiple-spin exchange interactions have not been 
discussed in the literature, although these  interactions may play an 
important role in some recently synthesized mixed-spin magnetic
materials and nanomagnets~\cite{kostyuchenko,schnack}. Here we  restrict 
ourselves to a general overview of the spin  phases of this interesting
mixed-spin model with cyclic interactions in the extreme quantum case
$(S_1,S_2)=(1,1/2)$. We also suppose  a FM effective 
exchange interactions  between the in-chain spins 
(i.e.,  $J_n\equiv J-3K/4 <0$). In the region $K>1.2J_1$,  
a detailed numerical ED analysis indicates at least two additional phases, 
denoted by DM and FM$_3$ in figure~\ref{phases}b~\cite{ivanov6}. 
DM is a dimerized  singlet phase stabilized  approximately in the region 
$1.5J_1\leq K\leq 2.3J_1$, whereas FM$_3$ is  a magnetic phase with a net FM 
moment. Finally, in the narrow interval $1.2J_1<K <1.5J_1$ a specific
ferrimagnetic  phase characterized by a  reduced magnetic moment per cell, 
$M< M_0=S_1-S_2$, seems to appear. Below, other mixed-spin models exhibiting 
similar exotic (non-Lieb-Mattis) ferrimagnetic phases are surveyed.
\subsection{Mixed-spin ladders with geometric frustration}
Some typical examples of mixed-spin  ladder structures
are shown in figure~\ref{models}.
The first two structures   reproduce, e.g.,
arrangements of the magnetic atoms   Mn
($S_1=5/2$) and Cu ($S_2=1/2$)
along the \textbf{a}-axis in the compounds
MnCu(pbaOH)(H$_2$O)$_3$ (pbaOH = 2-hydroxy-1,3-
propylenebisoxamato) and
MnCu(pba)(H$_2$O)$_3\cdot$2H$_2$O (pba =
1,3-propylenebisoxamato), respectively.
Along the \textbf{c}-axis, the magnetic ions
in both mixed-spin compounds are arranged as shown in
figure~\ref{models}(e)~\cite{kahn1}. 
A very recently synthesized quasi-1D mixed-valent-iron material 
[Fe$^{II}$Fe$^{III}$ 
(trans-1,4-cyclohexanedicarboxylate)$_{1.5}$]
exhibiting ferrimagnetic properties,  seems to be related to the frustrated 
mixed-spin  structure shown in figure~\ref{models}d~\cite{zheng1}. 
We are not aware of any real ferrimagnetic compound  related to the 
ladder structure shown in figure~\ref{models}e. Nevertheless, 
in view of the  easy control of the molecular-unit positions in the  
molecular chemistry, it may be expected that other materials, related to
the generic  mixed-spin models presented  in figure~\ref{models}, 
 will be synthesized in the near future.
\begin{figure}[htp]
\centering
\begin{minipage}[c]{0.4\textwidth}
\centering
\includegraphics[width=1\textwidth]{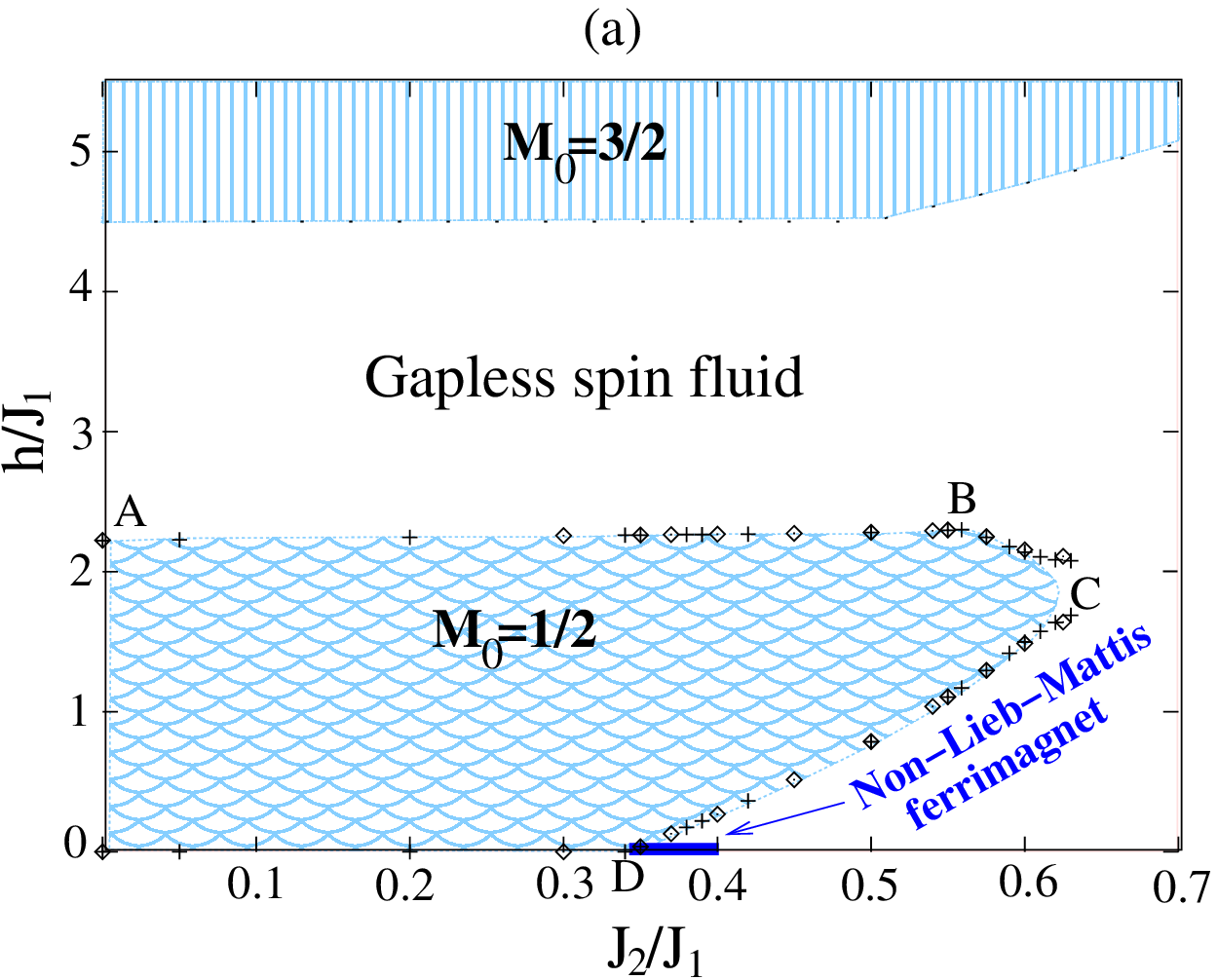}
\end{minipage}
\hspace{0.3in}
\begin{minipage}[c]{0.4\textwidth}
\centering
\includegraphics[width=0.9\textwidth]{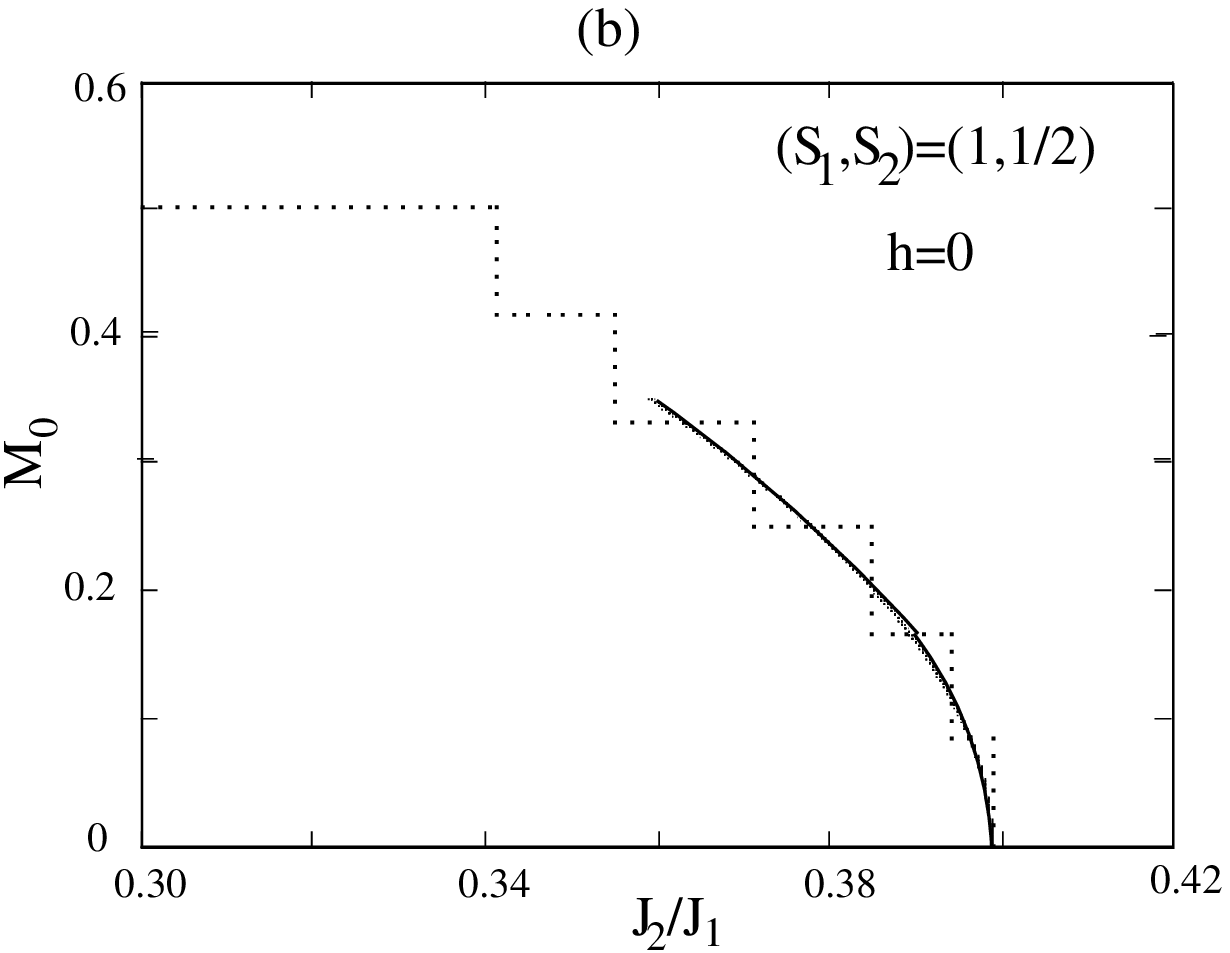}
\end{minipage}
\caption{(a) Phase diagram of the checkerboard mixed-spin $(1/1/2)$
ladder (figure~\ref{models}c) in the $(J_2/J_1,h/J_1)$ plane~\cite{ivanov8}.
The phase boundary of the fully polarized phase
($M_0=3/2$) is exact. Here $M_0$ is the FM moment per rung.
The boundary (ABCD) of the ferrimagnetic $M_0=1/2$ plateau phase
is obtained from numerical ED of periodic clusters with
$L=10$ (squares) and $L=12$ (crosses) rungs. The points B, C, and D
mark, respectively, the change of the lowest excited states, the tip of
the lobe, and the $h=0$ transition point at $J_2/J_1=0.399$ from  the  
$M_0<1/2$ ferrimagnetic phase to a gapless spin-fluid phases. 
The latter phase occupies the rest
of the phase diagram. 
(b) $M_0$ \textit{vs} $J_2/J_1$ for the
non-Lieb-Mattis (i.e.,  $M_0<1/2$) ferrimagnetic phase, 
as obtained from numerical ED of $L=12$
clusters (dashed line). The solid line connects the midpoints of the
stepse~\cite{ivanov9}. 
}
\label{ladder}
\end{figure}

On the theoretical side, unfrustrated variants of the ladder models
presented in figure~\ref{models} have already been analyzed in a number of
publication~\cite{senechal,fukui2,langari,ivanov7,trumper,aristov}. 
The unfrustrated checkerboard ladder ( figure~\ref{models}c) 
with AFM bonds ($J,J_1>0$) exhibits a ferrimagnetic ground state. 
Its low-energy 
properties closely reproduce the properties  of the generic 
mixed-spin chain discussed in Section 2~\cite{ivanov7}. Note that a variant
of this structure with FM legs ($J_1<0$) demonstrates completely different
features: The FM leg coupling drives the system into a
gapless  spin-fluid ground state which is  characteristic for the 
spin-1/2 AFM
Heisenberg chain~\cite{aristov}. Similar gapless phase appears in
the unfrustrated variant of the model displayed in 
figure~\ref{models}e~\cite{fukui2,langari},
whereas the model shown in figure~\ref{models}d possesses a gapped
non-degenerate rung-singlet ground state~\cite{trumper}, which is the
characteristic phase of the uniform  spin-1/2 Heisenberg ladder.

To understand the role of the geometric frustration in such
mixed-spin  models, let us turn to the  phase diagram in  figure~\ref{ladder}a  
presenting the phases of the frustrated $(1,1/2)$ checkerboard ladder 
(figure~\ref{models}c) in the parameter space 
$(J_2/J_1,h/J_1)$~\cite{ivanov8}. We are not aware of any publications
studying the other two frustrated models displayed in 
figures~\ref{models}d, and e.
 As a function of the frustration parameter $J_2/J_1$,
the classical phase diagram of the frustrated checkerboard model 
exhibits three phases, which can be described by the angles 
$(\theta,\phi)$ fixing the directions
of the classical spins $\mathbf{S}_{1,n}$ and  $\mathbf{S}_{2,n}$ 
in respect to the classical ferrimagnetic configuration  $(\theta,\phi)=(0,0)$
with up-$\mathbf{S}_{1,n}$ and down- $\mathbf{S}_{2,n}$ spins.
In the special case $(S_1,S_2)=(1,1/2)$, the classical canted state shown 
in figure~\ref{models}c  is stable in the interval
$0.3219J_1<J_2<0.4606J_1$. For larger $J_2$, a collinear
configuration with  $(\theta,\phi)=(\pi/2,\pi/2)$ is stabilized. 
In the classical limit,
the  transitions between the canted state and the other two phases are
continuous. Turning to the quantum model, the following changes in the
quantum phase diagram (line $h=0$ in figure~\ref{ladder}a) can be indicated.  
Whereas  the classical ferrimagnetic phase survives quantum fluctuations, 
the  collinear magnetic state is completely  destroyed.  
Instead, for $J_2>0.399J_1$ there appears a gapless
spin-fluid phase. In the general chase, the  quantum paramagnetic 
state is either critical (for half-integer $S_1+S_2$), or gapped 
(for integer  $S_1+S_2$).  

The most interesting changes appear in  the classical canted state.   
In figure~\ref{ladder}b we show numerical ED results for the FM moment per rung 
$M_0$ as function of the frustration parameter  $J_2/J_1$. We see that
the  quantum phase which substitutes the classical canted phase 
is characterized by a  finite FM moment which is, however, 
less that the quantized FM moment per rung ($S_1-S_2=1/2$) of a 
standard Lieb-Mattis ferrimagnet. Notice that this phase exists only at $h=0$. For
$h>0$ it merges into the Luttinger liquid phase (see figure~\ref{ladder}a). 
An extrapolation of the ED results for $L=8,10$ and $12$ rungs definitely
indicates the presence of this phase in the interval $0.341J_1<J_2<0.399J_1$. 
Already a qualitative  semiclassical analysis (supported by ED results)
implies that on  approaching the phase transition point at $J_2=0.341J_1$ 
from the ferrimagnetic phase, 
the lower magnon branch $E_k^{-}$ softens in the vicinity of  $k=\pi$, and
the gap at $k=\pi$ vanishes at the transition point to the canted phase.
Thus, there appears a linear  Goldstone mode which is  characteristic of 
the classical canted phase. It may be suggested that this critical mode 
survives quantum fluctuations~\cite{sachdev1}, whereas
the  spin rotation  symmetry $U(1)$ in the  $xy$ plane 
should be restored, i.e.,  
$\langle S_{1,n}^x\rangle =\langle S_{2,n}^x\rangle= 0$.
This scenario with a power-law decay of the transverse spin-spin
correlations is  supported by the renormalization-group analysis
of similar phases in quantum rotor models~\cite{sachdev1} as well as by
a recent DMRG analysis of a similar phase found in a generalized SDC model
with an additional competing AFM interaction between the off-chain spins
in figure~\ref{models}b~\cite{montenegro}. The first  studied 
quantum spin model exhibiting such an exotic quantum state seems to be  
the  spin-1/2 two-leg  ladder constructed from different 
(one FM and another AFM) legs  and AFM
rungs~\cite{tsukano}. Quite recently, there has been a number of reports
indicating  similar 1D quantum magnetic phases in some decorated quantum spin
chains~\cite{yoshikawa,hida} as well as in a mixed-spin $(2,1)$ Heisenberg
chain with  competing single-ion anisotropies~\cite{tonegawa2}.    
\section{Conclusion}
In conclusion, we have surveyed the available theoretical
results related to some  generic quantum  spin models displaying 
1D  ferrimagnetic ground states. The stress was  put on the interplay
between quantum fluctuations and the competing interactions.
Discussed  models of quasi-1D ferrimagnets with competing interactions 
exhibit a rich variety of magnetic and paramagnetic quantum  phases 
and provide unique examples of 1D  magnetic-paramagnetic quantum phase 
transitions. Finally, a  plenty of important open issues deserving further 
studies has been debated.

\section{Acknowledgements}
The author thanks the staff of the Fakult\"at f\"ur Physik, Universit\"at
Bielefeld for hospitality and  J\"urgen Schnack 
for the critical reading of the manuscript. This work was 
supported by   Deutsche Forschungsgemeinschaft (Grant SCHN 615/13-1)
and Bulgarian Science Foundation (Grant D002-264/18.12.08).

\end{document}